\pdfoutput=1

\documentclass[11pt,twoside,a4paper,cmspaper,final,collab]{cms-tdr}
\def\svnVersion{493286P}\def\cmsCernNoTag{CERN-EP-2018-272}\def\cmsCernDate{\today}\def\cmsMessage{Published in the Journal of High Energy Physics as \href{http://dx.doi.org/10.1007/JHEP03(2019)170}{\doi{10.1007/JHEP03(2019)170.}}}

\begin{document}\cmsNoteHeader{EXO-17-016}

\hyphenation{had-ron-i-za-tion}
\hyphenation{cal-or-i-me-ter}
\hyphenation{de-vices}
\RCS$HeadURL: svn+ssh://svn.cern.ch/reps/tdr2/papers/EXO-17-016/trunk/EXO-17-016.tex $
\RCS$Id: EXO-17-016.tex 490546 2019-02-28 17:27:28Z florez $
\newlength\cmsTabSkip\setlength{\cmsTabSkip}{1ex}

\providecommand{\PWR}{\ensuremath{\PW_{\cmsSymbolFace{R}}}\xspace}

\providecommand{\Nt}{\ensuremath{\cmsSymbolFace{N}_{\PGt}}\xspace}
\providecommand{\Ne}{\ensuremath{\cmsSymbolFace{N}_{\Pe}}\xspace}
\providecommand{\Nell}{\ensuremath{\cmsSymbolFace{N}_{\ell}}\xspace}
\providecommand{\Nmu}{\ensuremath{\cmsSymbolFace{N}_{\PGm}}\xspace}
\providecommand{\LQ}{\ensuremath{\mathrm{LQ}}\xspace}

\cmsNoteHeader{EXO-17-016}
\title{Search for heavy neutrinos and third-generation leptoquarks in hadronic states of two $\tau$ leptons and
two jets in proton-proton collisions at \texorpdfstring{$\sqrt{s}=13$\TeV}{sqrt(s) = 13 TeV}}

\date{\today}

\abstract{
A search for new particles has been conducted using events with two high transverse momentum $\tau$
leptons that decay hadronically and at least two energetic jets.
The analysis is performed using data from proton-proton collisions at $\sqrt{s}=13$\TeV, collected by the CMS experiment at the LHC in 2016 and corresponding to
an integrated luminosity of 35.9\fbinv. The observed data are consistent with standard model expectations.
The results are interpreted in the context of two physics models.
The first model involves right-handed charged bosons, $\PWR$, that decay to heavy right-handed Majorana neutrinos, $\Nell$ $(\ell= \Pe,\mu,\tau)$,
arising in a left-right symmetric extension of the standard model. The model considers that $\Ne$ and $\Nmu$ are too heavy to be detected at the LHC.
Assuming that the $\Nt$ mass is half of the $\PWR$ mass, masses of the $\PWR$ boson below 3.50\TeV
are excluded at 95\% confidence level. Exclusion limits are also presented considering different
scenarios for the mass ratio between $\Nt$ and $\PWR$, as a function of $\PWR$ mass.
In the second model, pair production of third-generation scalar leptoquarks that
decay into $\tau\tau\PQb\PQb$ is considered, resulting in an observed exclusion region with leptoquark masses below 1.02\TeV, assuming a 100\% branching
fraction for the leptoquark decay to a $\tau$ lepton and a bottom quark. These results represent the most stringent limits to date on these models.
}

\hypersetup{
pdfauthor={CMS Collaboration},
pdftitle={Search for heavy neutrinos and third-generation leptoquarks in final states with two hadronically decaying tau leptons and two jets in proton-proton collisions at $\sqrt{s}=13$\TeV},%
pdfsubject={CMS},
pdfkeywords={CMS, physics, leptoquarks, third generation}}

\maketitle
\section{Introduction}\label{sec:intro}

Despite its undeniable success, the standard model (SM) fails to answer some of the most fundamental questions in particle physics.
Among these are the source of matter-antimatter asymmetry, the particle nature of dark matter, the origin of dark energy, and the acquisition of neutrino mass.
The aim of this paper is to present a search for physics beyond the standard model in final states containing two hadronically
decaying $\tau$ leptons ($\tauh$) and two high transverse momentum (\pt) jets.
The analysis is performed using data from proton-proton ($\Pp\Pp$) collisions at $\sqrt{s}=13\TeV$, collected by the CMS experiment at the CERN LHC and
corresponding to an integrated luminosity of 35.9\fbinv.
To illustrate the sensitivity of this search for processes not included in the SM, two benchmark physics scenarios are considered for the interpretation
of the results: the production of heavy, right-handed Majorana neutrinos and the production of third-generation leptoquarks (\LQ{}s).
A description of the two models is given below.

\begin{figure}[ht]
 \centering
  \includegraphics[width=0.5\textwidth, height=0.25\textheight]{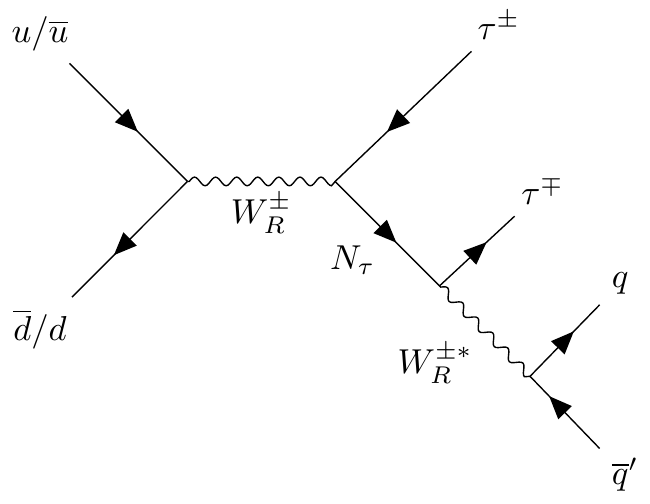}
  \caption{Leading order Feynman diagram for the production of a right-handed $\PWR$ that decays to a heavy neutrino
$\Nt$, with a final state of two $\tau$ leptons and two jets.}
\label{fig:feyn}
\end{figure}

The observation of neutrino oscillations implies nonzero neutrino masses, prompting a corresponding extension of the SM.
Results from neutrino oscillation experiments together with cosmological constraints imply very small values for these
masses~\cite{pdg, Lindner:2001hr, Minkowski:1977sc, Mohapatra:1979ia}.
The most popular explanation for very small neutrino masses is the ``seesaw" mechanism~\cite{GellMann:1980vs,Yanagida:1980xy,Mohapatra:1980qe}
in which the observed left-handed chiral states are paired with very heavy right-handed partners.
This mechanism can be realized in the left-right symmetric model
(LRSM)~\cite{Lindner:2001hr, Minkowski:1977sc, Mohapatra:1979ia}, in which the SM group SU(2)$_{\mathrm{L}}$ has a right-handed
counterpart, originally introduced to explain the nonconservation of parity in weak interactions.
The SU(2)$_{\mathrm{R}}$ group, similarly to SU(2)$_{\mathrm{L}}$, predicts the existence of three new gauge
bosons, $\cmsSymbolFace{W}_\cmsSymbolFace{R}^{\pm}$ and \PZpr, and three heavy right-handed Majorana neutrino states $\Nell$ ($\ell= \Pe,\mu,\tau$),
partners of the light neutrinos $\nu_{\ell}$.
A reference process allowed by this model is the production of a right-handed \PWR boson that decays to a heavy
neutrino and a lepton of the same generation ($\PWR \to \ell + \Nell \to \ell + (\ell\Pq\PAQq')$) and
gives rise to two jets and two leptons of the same flavor in the final state. Of particular interest for this analysis is the
scenario in which the $\PWR$ decay chain results in a pair of
high-\pt $\tau$ leptons, $\PWR \to \tau + \Nt \to \tau + (\tau\Pq\PAQq')$.
Figure~\ref{fig:feyn} shows the leading order (LO) Feynman diagram for the production of a $\Nt$.

\begin{figure}[ht]
\centering
    \includegraphics[width=0.45\textwidth]{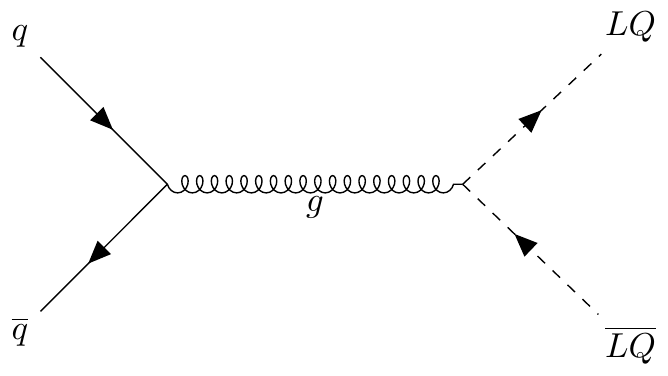}
    \includegraphics[width=0.45\textwidth]{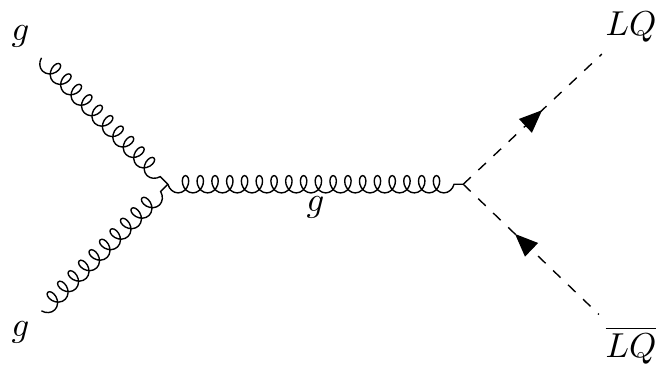}
    \includegraphics[width=0.45\textwidth]{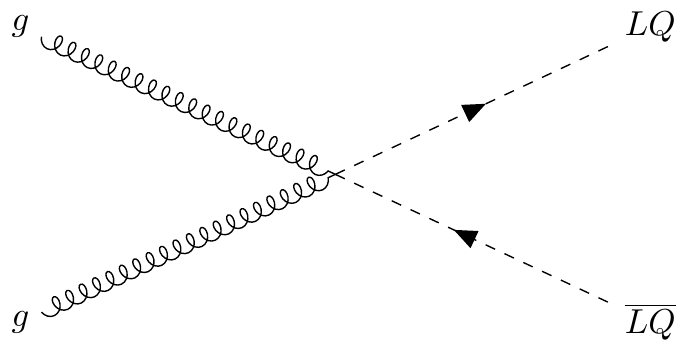}
    \includegraphics[width=0.45\textwidth]{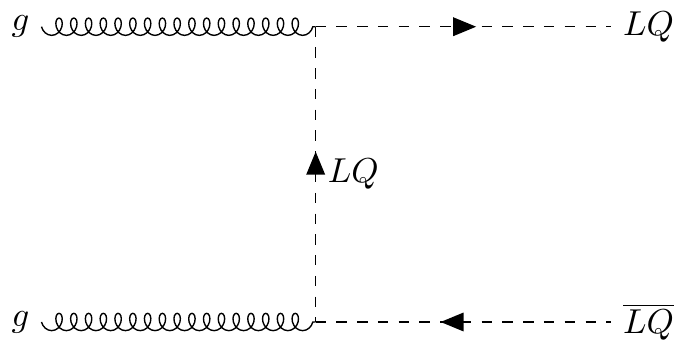}
  \caption{Leading order Feynman diagrams for the pair-production of \LQ{}s, leading to final states with two $\tau$ leptons and two \PQb quarks.}
\label{fig:feyn2}
\end{figure}

A similar $\tau\tau \mathrm{jj}$ final state can be realized in other extensions of the SM, such as grand unified
theories~\cite{Pati:1973uk, PatiSalam, GeorgiGlashow, Fritzsch:1974nn},
technicolor models~\cite{Dimopoulos:1979es,Dimopoulos:1979sp,Technicolor, Lane:1991qh},
compositeness scenarios~\cite{LightLeptoquarks,Gripaios:2009dq},
and $R$ parity~\cite{Farrar:1978xj} violating supersymmetry~\cite{Ramond:1971gb,Golfand:1971iw,Neveu:1971rx,Volkov:1972jx,Wess:1973kz,Wess:1974tw,Fayet:1974pd,Nilles:1983ge,Barbier:2004ez}.
These theories predict a new scalar or vector boson, referred to as a leptoquark in the literature, which carries nonzero lepton and baryon numbers,
as well as color and fractional electric charge \cite{PatiSalam,Gripaios:2009dq}.
In order to comply with experimental constraints on flavor changing neutral currents and other rare processes \cite{Buchmuller:1986zs,Shanker:1982nd},
three types of \LQ{}s are generally considered, each coupled to the leptons and quarks of its generation.
The \LQ{}s recently gained notable theoretical attention as one of the most suitable candidates to explain
the  $\PaB\to\PD^{*}\tau\nu$ and $\PQb\to\PQs\ell\ell$ anomalies reported by the BaBar~\cite{Lees:2012xj,Lees:2013uzd}, Belle~\cite{Matyja:2007kt,Bozek:2010xy,Huschle:2015rga,Hirose:2016wfn},
and LHCb~\cite{Aaij:2013qta,Aaij:2014ora,Aaij:2015yra,Aaij:2015oid,Aaij:2017vbb} Collaborations. In particular, models containing enhanced
couplings to the third-generation SM particles are favored to interpret these results \cite{Tanaka:2012nw, Sakaki:2013bfa, Dorsner:2013tla, Gripaios:2014tna}.
In this search, we consider pair-produced scalar \LQ{}s, each decaying to a $\PGt$ lepton and a bottom
quark (\PQb). Figure~\ref{fig:feyn2} shows the LO Feynman diagrams for the pair-production of \LQ{}s.

The most recent heavy neutrino and \LQ searches in $\ell\ell\mathrm{jj}$ final states
have been carried out by the
ATLAS~\cite{ATLAS:2012ak,Aad:2015xaa,ATLAS:2013oea} and CMS~\cite{Wr2016,Sirunyan:2017xnz,Sirunyan:2017yrk,Khachatryan:2016jqo,Khachatryan:2014dka} Collaborations.
The most stringent limits in the $\tau\tau\mathrm{jj}$ final states are set in Ref.~\cite{Sirunyan:2017yrk} and exclude \PWR masses below 2.9\TeV,
assuming that the mass of the right-handed neutrino is half of the mass of the \PWR boson, and scalar \LQ masses below
850\GeV, assuming that the \LQ decays to a $\tau$
lepton and a bottom quark with 100\% branching fraction.
Moreover, searches for third-generation \LQ{}s have been performed in other final states: pairs of scalar \LQ{}s each of which
decays to a $\tau$ lepton and a top quark \cite{Sirunyan:2018nkj}, pairs of scalar and vector \LQ{}s each of which decays to a
quark (top, bottom, or light-flavor) and a neutrino \cite{Sirunyan:2018kzh}, and singly produced scalar \LQ{}s in association with a $\tau$
lepton with the \LQ decaying to a $\tau$ lepton and a bottom quark~\cite{Sirunyan:2018jdk}.
In this analysis we focus on the $\tau\tau\mathrm{jj}$ search channel in which both of the $\tau$ leptons decay hadronically.
Hadronic $\tau$ lepton decays account for approximately 65\% of all possible $\tau$ lepton final states, so that the pair
branching fraction is $42$\%.

The paper is organized as follows. Section~\ref{sec:cmsDetector} gives a brief description of the CMS detector.
The event reconstruction is described in Section~\ref{sec:leptonRecoId}, followed by the description  of the simulation of the signal and background samples in Section~\ref{sec:backgrounds}.
The selection criteria defining the signal region (SR), described in Section~\ref{sec:eventSelections}, reduce the background contributions to achieve maximum discovery potential.
A main challenge of this analysis is to achieve high and well-understood signal selection and trigger efficiencies, with small systematic uncertainty,
with SM signatures containing genuine $\tauh$ candidates. The strategy is described in Section~\ref{sec:estimation} and
relies on the selection of $\PZ(\to \ell\ell$)+jets events.
A number of additional background-enriched regions are described in Section~\ref{sec:estimation}.
These regions are defined to minimize the systematic uncertainty of the background contributions as well as to cross-check the accuracy of the
efficiency measurements. Relevant systematic uncertainties are described in Section~\ref{sec:systematics}.
The results are presented in Section~\ref{sec:Results}. The paper concludes with a summary in Section~\ref{sec:Summary}.

\section{The CMS detector}\label{sec:cmsDetector}

A detailed description of the CMS detector, together with a definition of the coordinate system
used and the relevant kinematic variables, can be found in \cite{CMS}.
The central feature of the CMS apparatus is a
superconducting solenoid of 6\unit{m} inner diameter, providing a field
of 3.8\unit{T}. Within the field volume are the silicon pixel and strip
tracker, the crystal electromagnetic calorimeter (ECAL),
which includes a silicon sensor preshower detector in front of the ECAL endcaps,
and the brass and scintillator hadron calorimeter. Muons are measured in
gas-ionization detectors embedded in the steel return yoke. In
addition to the barrel and endcap detectors, CMS has extensive forward
calorimetry. The inner tracker measures charged particles within pseudorapidity range $\abs{\eta} < 2.5$
and provides an impact parameter resolution of $\sim$15\micron and a transverse momentum
resolution of about 1.5\% for 100\GeV particles.
Collision events of interest are selected using a two-tiered trigger system.
The first level, composed of custom hardware processors,
selects events at a rate of around 100\unit{kHz}.
The second level, based on an array of microprocessors
running a version of the full event reconstruction software
optimized for fast processing,
reduces the event rate to around 1\unit{kHz} before data storage.

\section{Event reconstruction and particle identification}\label{sec:leptonRecoId}

Jets are reconstructed using the particle-flow (PF) algorithm~\cite{CMS-PRF-14-001}.
In the PF approach, information from all detectors is combined to reconstruct and identify final-state particles
(muons, electrons, photons, and charged and neutral hadrons) produced in the $\Pp\Pp$ collision.
PF particles are clustered into jets using the anti-\kt clustering algorithm~\cite{Cacciari:2008gp} with a distance parameter of $0.4$.
Jets are required to pass identification criteria designed to reject
anomalous behavior from the calorimeters.
The identification efficiency is $>$99\% for jets with $\pt>30\GeV$ and $\abs{\eta}<$ 2.4 that are within the tracking
acceptance~\cite{CMS-PAS-JME-14-003}.
The jet energy scale and resolution in simulation are corrected to match their measured values in data using factors that depend on the
\pt and $\eta$ of the jet \cite{CMS:JetResol,CMS-PAS-JME-14-001}.
Jets originating from the hadronization of bottom quarks are identified using the combined secondary vertex algorithm \cite{CMSbjet2}
which exploits observables related to the long lifetime of \PQb hadrons.
For \PQb quark jets with $\pt>30\GeV$ and $\abs{\eta}<2.4$, the algorithm's identification efficiency at the loose working point used in
this analysis is about $80\%$, while misidentification rate for light-quark and gluon jets is about $10\%$~\cite{CMSbjet2}.
Although a \PQb-tagged jet requirement is not used to define the \LQ SR, \PQb quark jets are
used to obtain \ttbar-enriched control samples for estimation of the background rate in the SR.

Although muons and electrons are not used to define the SR, they are utilized to obtain control samples for the background estimations.
Electron candidates are reconstructed by first matching clusters of energy deposited in the
ECAL to reconstructed tracks. Selection criteria based on the distribution of the shower shape,
track-cluster geometric matching, and consistency between the cluster energy and track momentum are
then used in the identification of electron candidates~\cite{Khachatryan:2015hwa}.
Muons are reconstructed using the tracker and muon chambers.
Quality requirements based on the minimum number of measurements in the silicon
tracker, pixel detector, and muon chambers are applied to suppress
backgrounds from decays in flight and hadron shower remnants that reach the muon system \cite{Sirunyan:2018fpa}.
The muon and electron identification efficiencies for the quality requirements and kinematic range used in this analysis are larger than $98\%$.

The electron and muon candidates are required to satisfy isolation criteria in order to reject
nonprompt leptons that originate from the hadronization process.
Isolation is defined as the scalar sum of the \pt values of reconstructed charged and neutral particles within a cone of radius $\Delta R =
\sqrt{\smash[b]{(\Delta\eta)^{2} + (\Delta\phi)^{2}}}=0.4$ around
the lepton-candidate track, excluding the lepton candidate, divided by the \pt of the lepton candidate.
A correction is applied to the isolation variable to account for the effects of additional $\Pp\Pp$ interactions (pileup)~\cite{Cacciari:2007fd}.
For charged particles, only tracks associated with the primary vertex are included in the
isolation sums.
The reconstructed vertex with the largest value of summed physics-object $\pt^2$ is taken to be the primary $\Pp\Pp$ interaction vertex.
The corresponding physics-objects are the leptons, jets, and the missing transverse momentum (\ptmiss) reconstructed from those objects.
The jets are clustered using the anti-\kt jet finding algorithm~\cite{Cacciari:2008gp,Cacciari:2011ma} with the tracks assigned to the vertex as inputs.

Hadronic decays of the $\tau$ lepton are reconstructed and identified using the hadrons-plus-strips algorithm \cite{Sirunyan:2018pgf},
designed to optimize the performance of $\tauh$ reconstruction by considering specific $\tauh$
decay modes. This algorithm starts from anti-\kt jets and reconstructs $\tauh$ candidates from tracks (also referred to as ``prongs'') and
energy deposits in strips of the ECAL, in the 1-prong, 1-prong + $\pi^{0}$, 2-prong, and 3-prong decay modes. The 2-prong decay mode allows $\tauh$
candidates to be reconstructed even if one track has not been reconstructed. However, given the large rate for jets to be misidentified in this decay
mode, the 2-prong decay mode is not used to reconstruct $\tauh$ candidates in the signal region of this analysis.
To suppress backgrounds from light-quark or gluon jets, identification and isolation conditions are enforced by requiring
the $\tauh$ candidates to pass a threshold value of a multivariate (MVA) discriminator \cite{Sirunyan:2018pgf}
that takes isolation variables and variables related to the $\tau$ lepton lifetime as input.
The isolation variables are calculated using a cone of radius $\Delta R=0.5$ in the vicinity of the identified $\tauh$ candidate and
considering the energy deposits of particles not included in the reconstruction of the $\tauh$ decay mode.
The ``tight" MVA isolation working point \cite{Sirunyan:2018pgf} is used to define the SR, which results in a $\tauh$
identification efficiency of typically $55\%$ for the kinematic range used in this analysis.
Additionally, $\tauh$ candidates are required to be distinguishable from electrons and muons.
The algorithm to discriminate a $\tauh$ from an electron utilizes observables that quantify the compactness and shape of energy deposits in the ECAL, to
distinguish electromagnetic from hadronic showers, in combination with observables that are sensitive to the amount of bremsstrahlung
emitted along the leading track and to the overall particle multiplicity. The discriminator against muons
is based on the presence of measurements in the muon system associated with the track of the $\tauh$ candidate.

The presence of neutrinos from the $\tau\tau$ decays must be inferred from the imbalance of total momentum in the detector.
The magnitude of the negative vector sum of the transverse momenta of visible PF objects is the missing transverse momentum.
Information from the forward calorimeter is included in the calculation of \ptmiss, and the
jet corrections described above are propagated as corrections to \ptmiss~\cite{CMS-PAS-JME-17-001}.
Missing transverse momentum is one of the most important observables for differentiating the signal events from background events that do not contain neutrinos, such as quantum
chromodynamics (QCD) multijet events.

\section{Signal and background samples}\label{sec:backgrounds}

The production of top quark pairs (\ttbar), the production of a \PZ boson decaying to a $\tauh$ pair plus
associated jets from initial-state radiation ({\PZ}+jets), and QCD multijet processes are the prevailing backgrounds for this search.
Background from \ttbar events is characterized by two \PQb quark jets in addition to
genuine isolated $\tauh$ leptons.
The contribution of {\PZ}+jets events constitutes an irreducible background since it has the same final state containing genuine,
well-isolated $\tauh$ candidates, associated energetic jets, and true \ptmiss from neutrinos present in the
$\tau$ lepton decays. The QCD multijet events are characterized by jets with a high-multiplicity of particles, which can be misidentified
as $\tauh$.

To estimate the main backgrounds, a combination of Monte Carlo (MC) simulated samples and techniques based on data are employed.
The dominant backgrounds are estimated from data, using control regions (CR) enriched in the contributions of targeted background processes and with negligible contamination from signal events.
Samples of events produced by MC simulation are used to extrapolate background yields from a CR to the SR and to model the shape of
the of the distributions of observables defined in Sec.~\ref{sec:eventSelections} aiming to estimate the mass of the $\PWR$
($m(\tau_{\mathrm{h},1},\tau_{\mathrm{h},2},\mathrm{j}_{1},\mathrm{j}_{2},\ptmiss)$) and that of the \LQ ($S^{\mathrm{MET}}_{\mathrm{T}}$).
Subdominant background contributions are estimated using MC simulations.
The \MGvATNLO 2.6.0 program \cite{aMCatNLO} is used for {\PZ}+jets, {\PW}+jets, {\ttbar}+jets,
and single-top quark production. The \MGvATNLO generator is interfaced with {\PYTHIA8.212}~\cite{Sjostrand:2014zea}, using the CUETP8M1 tune~\cite{Khachatryan:2015pea}, for parton shower
and fragmentation. The LO {\PYTHIA} generator is used to model the diboson (VV) processes.
The MC background and signal yields are normalized to the integrated luminosity using next-to-next-to-leading
order or next-to-leading order (NLO) cross sections \cite{Kramer:2004df}.

The $\Nt$ signal samples are generated at the leading order using {\PYTHIA8.212}
with $\PWR$ masses ranging from 1 to 4\TeV, in steps of 0.25\TeV.
It is assumed that the gauge couplings associated with the left- and right-handed SU(2) groups are equal and the $\Nt$ decays are prompt.
It is also assumed that the $\Ne$ and $\Nmu$ are too heavy to play a role in the decay of
$\PWR$, and thus $\PWR \to \tau \Nt$ and $\PWR \to \Pq \PAQq'$
are the dominant decay modes. The branching fraction for the $\PWR \to \tau \Nt$ decay is approximately 10--15\%,
depending on the $\PWR$ and $\Nt$ masses.
For the $\PWR$ mass range of interest for this analysis,
the $\Nt \to \tau \Pq \PAQq'$ branching fraction is close to 100\%.
The signal cross sections are calculated at the NLO accuracy. The ratios of the NLO and the LO results provide factors of 1.3,
known as $K$ factors, for the $\PWR$ mass range relevant to this analysis~\cite{delAguila:2007qnc}.

Simulated samples for the scalar \LQ signal processes are generated
for a range of masses between 250 and 1500\GeV in steps of
50\GeV. The signal MC generation uses {\PYTHIA8.212} and CTEQ6L1 parton
distribution functions (PDF)~\cite{Pumplin:2002vw}.
Signal cross sections are calculated at NLO
accuracy using the CTEQ6.6M PDF set~\cite{Kramer:2004df}.
The NLO-to-LO $K$ factors range from 1.3 to $2.0$ in the mass range 200--1500\GeV~\cite{Kramer:2004df}.
The branching fraction of the \LQ to a $\tau$ lepton and a \PQb quark is
assumed to be 100\%.

The mean number of interactions in a single bunch crossing in the analysed dataset is 23.
In MC events, multiple interactions are superimposed on the primary collision, and each MC event
is re-weighted such that the distribution of the number of true interactions matches that in data.

\section{Event selection}\label{sec:eventSelections}

Events are selected with a trigger requiring at least two $\tauh$ candidates with $\pt > 32\GeV$ and $\abs{\eta} < 2.1$ \cite{Sirunyan:2018pgf}.
Additional kinematic criteria on \pt and $\eta$ are applied to achieve a trigger efficiency greater than $90\%$
per $\tauh$ candidate.
Preselected events are required to have at least two $\tauh$ candidates, each with $\pt > 70\GeV$ and
$\abs{\eta} < 2.1$. The $\abs{\eta} < 2.1$ requirement ensures that the $\tauh$ candidates are fully reconstructed within the tracking acceptance.
In addition, the two $\tauh$ candidates must be separated by $\DR > 0.4$, to avoid overlaps.
Selected $\tauh$ candidates must also pass the reconstruction and identification criteria described in Section~\ref{sec:leptonRecoId}.
In the LRSM, $\tau\tau$ pairs can be of the opposite or same-sign charge. 

The associated jet selection criteria include at least two jets with $\pt > 50\GeV$ and $\abs{\eta} < 2.4$.
To avoid overlaps, only jet candidates separated from the selected $\tauh$ candidates by $\Delta R > 0.4$ are considered.
The background contribution from QCD multijet events is larger in this analysis than in channels with one or both $\tau$ leptons decaying leptonically.
To suppress the contribution from QCD multijet events, \ptmiss is required to be larger than 50\GeV.
Finally, the visible invariant mass of the $\tauh\tauh$ pair, $m(\tau_{\mathrm{h},1},\tau_{\mathrm{h},2})$, is chosen to be greater than 100\GeV, to reduce
the {\PZ}+jets contribution.

The visible $\tau$ lepton decay products, the two highest \pt jets, and the missing transverse momentum vector \ptvecmiss are used to define an
observable for each benchmark scenario considered in the analysis.
The heavy neutrino search strategy consists in looking for a broad enhancement of events above the expected background in the distribution of the partial mass indicative of
new physics, defined as:
\begin{equation*}
m(\tau_{\mathrm{h},1},\tau_{\mathrm{h},2},\mathrm{j}_{1},\mathrm{j}_{2},\ptmiss) = \sqrt{(E_{\tau_{\mathrm{h},1}}+E_{\tau_{\mathrm{h},2}}+E_{\mathrm{j}_{1}}+E_{\mathrm{j}_{2}}+\ptmiss)^{2}-(\vec{p}_{\tau_{\mathrm{h},1}}+\vec{p}_{\tau_{\mathrm{h},2}}+\vec{p}_{\mathrm{j}_{1}}+\vec{p}_{\mathrm{j}_{2}}+\ptvecmiss)^{2}}.
\end{equation*}
On average the partial mass is large in the heavy-neutrino case, $\langle m(\tau_{\mathrm{h},1},\tau_{\mathrm{h},2},\mathrm{j}_{1},\mathrm{j}_{2},\ptmiss) \rangle \approx m(\PWR)$.
For the pair production of \LQ{}s, the scalar sum of the transverse momenta of the decay products and the \ptmiss,
$S^{\mathrm{MET}}_{\mathrm{T}}=\pt^{\tau_{\mathrm{h},1}}+\pt^{\tau_{\mathrm{h},2}}+\pt^{j_{1}}+\pt^{j_{2}}+\ptmiss$, is expected to be large ($\langle S^{\mathrm{MET}}_{\mathrm{T}} \rangle \approx m(\LQ)$).
The analysis explores the possibility of an excess of events with respect to the background prediction
in the upper range of the $S^{\mathrm{MET}}_{\mathrm{T}}$ distribution.
The $S^{\mathrm{MET}}_{\mathrm{T}}$ variable provides better significance in comparison to the
$S_{\mathrm{T}}=\pt^{\tau_{\mathrm{h},1}}+\pt^{\tau_{\mathrm{h},2}}+\pt^{j_{1}}+\pt^{j_{2}}$ variable used in the prior \LQ search in the
$\tauh\tauh\mathrm{jj}$ channel~\cite{Khachatryan:2016jqo}.

The set of events satisfying the preselection together with the associated jet selection define the SR.
The total expected background yield in the SR, estimated from simulation, is $126$ events, with
\ttbar, QCD multijet, {\PZ}+jets, {\PW}+jets, single-top quark, and diboson production
composing 38.0, 27.0, 18.4, 11.0, 4.0 and, 1.6\% of the rate, respectively.
The analysis strategy is similar to that of previous heavy neutrino and leptoquark searches~\cite{Khachatryan:2016jqo,Sirunyan:2017yrk}.
However, unlike heavy neutrino searches in the $\Pe\Pe\mathrm{jj}$ or $\mu\mu \mathrm{jj}$ final states~\cite{Khachatryan:2014dka, ATLAS:2012ak},
the $\PWR$ resonance mass in the $\tauh\tauh\mathrm{jj}$ channel cannot be fully reconstructed because
of the presence of neutrinos from the $\tau$ lepton decays.

The signal selection efficiency for the $\PWR$ process, assuming that the $\Nt$ mass is half of the
$\PWR$ mass, is 2.0\% for $m(\PWR)=1.0\TeV$ and 6.6\% for $m(\PWR)=4.0\TeV$.
The corresponding efficiency for $\LQ\to\tau \PQb$ events is 5.1\% for $m(\LQ)=0.6\TeV$ and 8.2\% for
$m(\LQ)=1.0\TeV$. These efficiencies include the 42\% branching fraction of $\tau\tau$ to $\tauh\tauh$.

\section{Background estimation}\label{sec:estimation}

The \ttbar, QCD multijet, and {\PZ}+jets processes are expected to account for $84 \%$ of the total background.
Dedicated CRs are used to check the modeling of \ttbar and {\PZ}+jets
events in simulation and to determine if any corrections need be applied.
The estimation of the QCD multijet background is performed using a method fully based on data.
The remaining contributions arising from {\PW}+jets, single-top quark, and diboson events are obtained from simulation.

A \ttbar-enriched control sample is obtained with similar selections to the SR, except selecting two well-identified muons
instead of two $\tauh$ candidates, requiring at least one \PQb-tagged jet, and vetoing dimuon candidates around the
\PZ boson mass peak ($80 < m_{\mu\mu} < 110\GeV$). Since the dijet and \ptmiss selection criteria are the same
as in the SR, the data-to-simulation scale factor $SF^{\ttbar}_{\mu\mu}=0.93 \pm 0.01$ measured in this CR represents a correction for the modeling of the dijet
and \ptmiss selection efficiencies by simulation.

\begin{figure}
 \centering
  \includegraphics[width=0.48\textwidth, height=0.35\textheight]{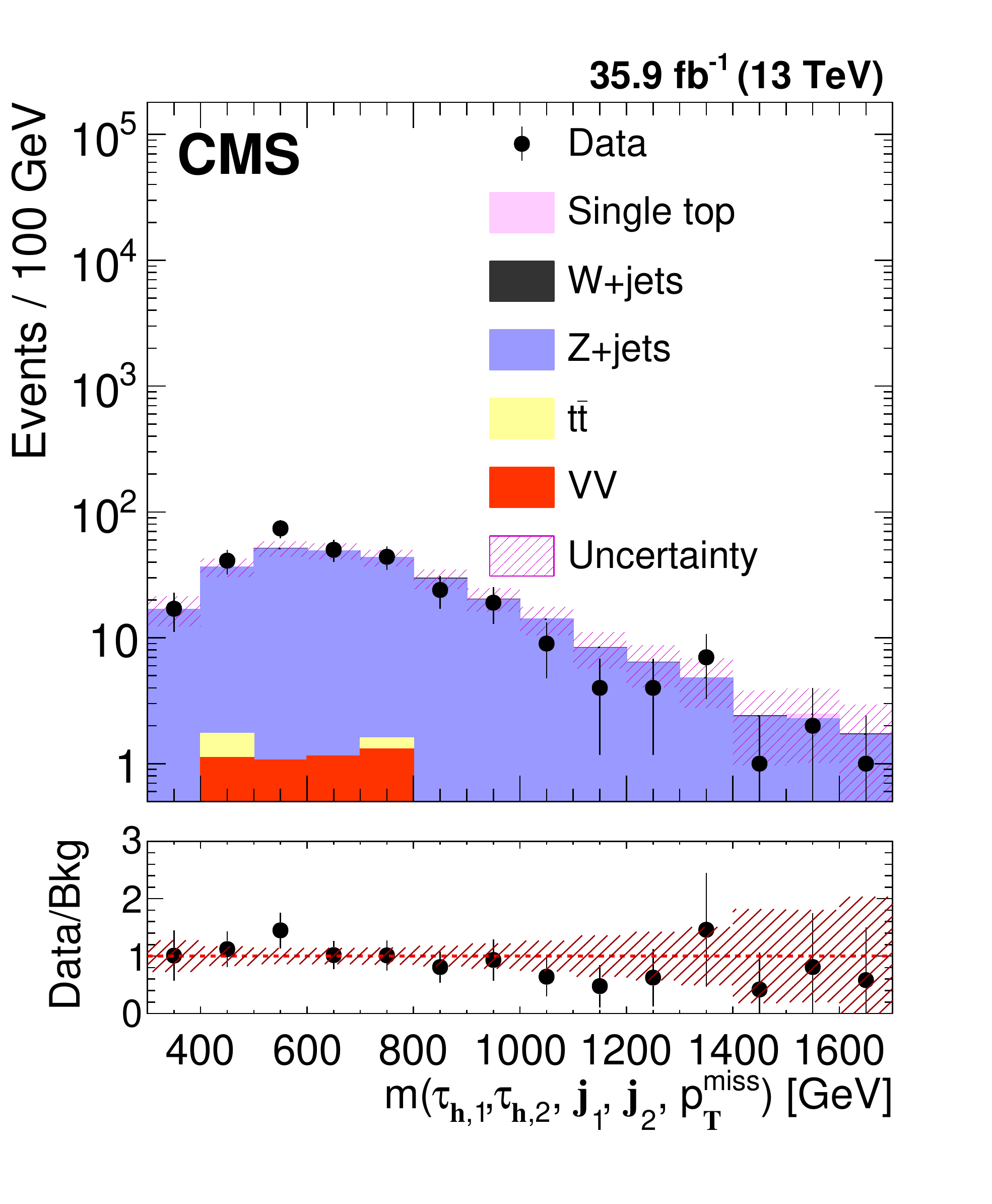}
  \includegraphics[width=0.48\textwidth, height=0.35\textheight]{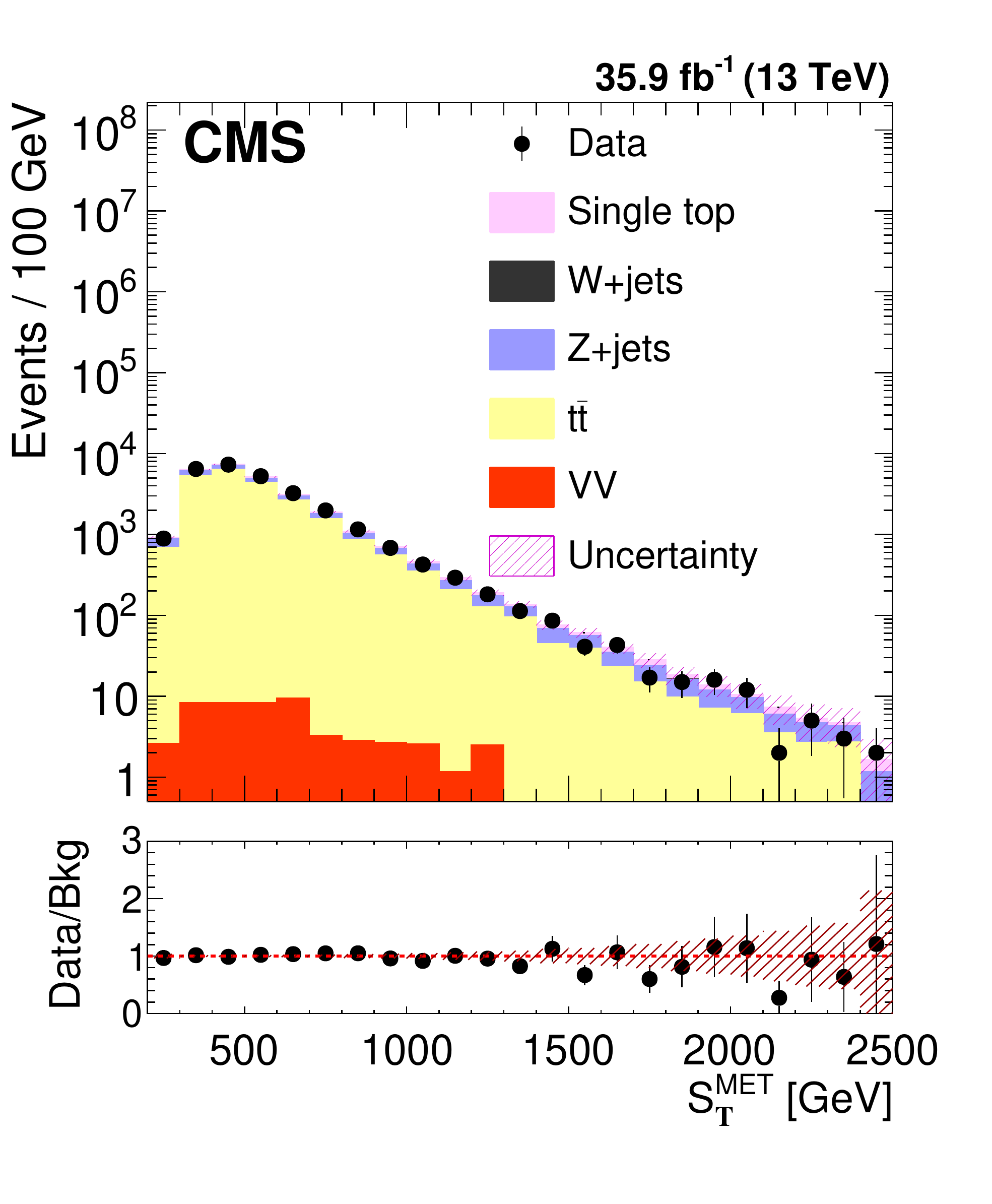}
  \caption{Distributions in $m(\tau_{\mathrm{h},1},\tau_{\mathrm{h},2},\mathrm{j}_{1}, \mathrm{j}_{2},\ptmiss)$ (left), for the
$\PZ (\tau\tau)$ control sample with relaxed $\tauh$ candidate \pt thresholds and $m(\tau_{\mathrm{h},1},\tau_{\mathrm{h},2}) < 100\GeV$,
and $S^{\mathrm{MET}}_{\mathrm{T}}$ (right), for the $\ttbar (\mu \mu \mathrm{jj})$ control sample.
The bottom frames show the ratio between the observed data in the control samples and the total background (Bkg) predictions. The bands correspond to the statistical uncertainty for the
background.}
\label{fig:TTmumuCR}
\end{figure}

Figure~\ref{fig:TTmumuCR} (right) shows the $S^{\mathrm{MET}}_{\mathrm{T}}$ distribution in this CR, after correcting the \ttbar normalization from simulation using the measured scale
factor $SF^{\ttbar}_{\mu\mu}$.
The agreement gives confidence that the $S^{\mathrm{MET}}_{\mathrm{T}}$ shape for the \ttbar background can be taken from simulation.
An alternate estimate of the scale factor is obtained from a CR defined with the same dijet and
\ptmiss requirements as for the SR but selecting events with one muon and one electron
(instead of a $\tauh\tauh$ pair).
The resulting estimate, $SF^{\ttbar}_{\Pe \mu}=0.90 \pm 0.01$, is combined with the measurement from the dimuon CR; the average of the two scale factors
($SF^{\ttbar}$) is used to estimate the \ttbar
prediction in the SR, and the absolute difference between the two scale factors, 3\%, is considered a systematic uncertainty in the estimated \ttbar yield.
Therefore, the \ttbar
contribution in the SR, $N^{\ttbar}_{\mathrm{SR}}$, is given by
$N^{\ttbar}_{\mathrm{SR}}=N^{\ttbar}_{\mathrm{SR}}(\mathrm{MC}) SF^{\ttbar}$.

The measurement of the {\PZ}+jets background component is based on both simulation and data.
Ideally the {\PZ}+jets contribution to the SR would be obtained using a CR obtained with similar $\tauh\tauh\mathrm{jj}$ criteria
to the SR, but with minimal modifications to the selection to achieve negligible signal contamination. However, such a CR has too few events,
resulting in large systematic uncertainty. Instead, since the efficiency of the requirement of two high quality $\tauh$ candidates
is known to be well modeled by simulation~\cite{Sirunyan:2018pgf}, we use a {\PZ}+jets-enriched control sample
obtained by requiring two well-identified muons with an invariant mass compatible with the \PZ-mass peak, instead of two $\tauh$ candidates, and all of the other event selection
criteria used in the SR. Since muons are produced in \PZ-decays as often as $\tau$ leptons, a $\mu\mu\mathrm{jj}$ control sample can be used to measure a correction factor
$SF^{\PZ\to\mu\mu}_{\text{dijet}}$ for the modeling of two additional jets, independently from the $\tauh\tauh$
requirement, and with reduced systematic uncertainty. Candidate events for the $\PZ(\to\mu\mu)$+jets control sample were collected using a trigger that requires
at least one isolated muon with $\pt(\mu) > 24\GeV$ per event. The measured correction factor is $SF^{\PZ\to\mu\mu}_{\text{dijet}}=1.02 \pm 0.02$.
Therefore, the $\PZ(\to\tau\tau)$ contribution in the SR can be calculated as $N^{\PZ\to\tau\tau}_{\mathrm{SR}}=N^{\PZ\to\tau\tau}_{\mathrm{SR}}(\mathrm{MC}) SF^{\PZ\to\mu\mu}_{\text{dijet}}$.
The modeling of the shapes of the $m(\tau_{\mathrm{h},1},\tau_{\mathrm{h},2},\mathrm{j}_{1},\mathrm{j}_{2},\ptmiss)$ and
$S_{\mathrm{T}}^{\mathrm{MET}}$ distributions is checked in $\PZ(\to\tau\tau)$+jets events that pass relaxed conditions on the $\tauh$ \pt threshold ($\pt > 60\GeV$) and an inverted requirement on
the mass of the $\tauh\tauh$ pair ($m(\tau_{\mathrm{h},1},\tau_{\mathrm{h},2})<100\GeV$).
Figure~\ref{fig:TTmumuCR} (left) shows the $m(\tau_{\mathrm{h},1},\tau_{\mathrm{h},2},\mathrm{j},\ptmiss)$ distribution
in this CR. The simulated and observed distributions of $m(\tau_{\mathrm{h},1},\tau_{\mathrm{h},2},\mathrm{j}_{1},\mathrm{j}_{2},\ptmiss)$
and $S_{\mathrm{T}}^{\mathrm{MET}}$ are found to be in agreement.

Events from QCD multijet processes become a background when two jets are misidentified as $\tauh$ candidates.
To avoid reliance on simulation, which may not be trustworthy at the high values of \pt, $m(\tau_{\mathrm{h},1},\tau_{\mathrm{h},2},\mathrm{j}_{1},\mathrm{j}_{2},\ptmiss)$,
and $S_{\mathrm{T}}^{\mathrm{MET}}$ of the search region, the QCD multijet background is estimated from data using
the matrix (``$ABCD$'') method.
Since \ptmiss and $\tauh$ isolation are the main discriminating variables against QCD multijet events,
the estimation methodology for this background utilizes CRs obtained by inverting the requirements on these observables.
It has been checked that the \ptmiss and the $\tauh$ isolation variables are uncorrelated.
In the remainder of this section, events obtained by inverting the isolation requirement on both $\tauh$ candidates will be referred
to as nonisolated $\tauh\tauh$ samples. The regions used to perform the
QCD multijet estimation, referred to as $ABCD$, are defined as follows:

\begin{itemize}
  \item A: $\ptmiss < 50\GeV$; fail the tight but pass the loose $\tauh$ isolation
  \item B: $\ptmiss < 50\GeV$; pass the tight $\tauh$ isolation
  \item C: $\ptmiss > 50\GeV$; fail the tight but pass the loose $\tauh$ isolation
  \item D: $\ptmiss > 50\GeV$; pass the tight $\tauh$ isolation
\end{itemize}

Note that region $D$ corresponds to the SR. The regions $A$, $B$ and $C$ are enriched in QCD multijet events (78--96\% depending on the region).
We estimate the QCD multijet component in the SR as
$N_{\mathrm{QCD}}^{D} = N_{\mathrm{QCD}}^{C} ({N_{\mathrm{QCD}}^{B}}/{N_{\mathrm{QCD}}^{A}})$, where
contributions from non-QCD backgrounds ($N_{\neq \mathrm{QCD}}$) are subtracted from data in each region $i=A,B,C$ using the MC prediction
($N_{\mathrm{QCD}}^{i}=N_{\mathrm{Data}}^{i}-N_{\neq \mathrm{QCD}}^{i}$).
Here ${N_{\mathrm{QCD}}^{B}}/{N_{\mathrm{QCD}}^{A}}$ is referred to as the isolation ``tight-to-loose" (TL) ratio.
The shapes of QCD multijet events in data containing two nonisolated $\tauh$ candidates are normalized using the TL ratio.
This procedure yields a QCD multijet estimate of $N_{\mathrm{QCD}}^{\mathrm{SR}} = 33.8 \pm 6.0$. The uncertainty is based on the
event counts in the data and MC samples.

\begin{figure}\centering
  \includegraphics[width=0.48\textwidth, height=0.35\textheight]{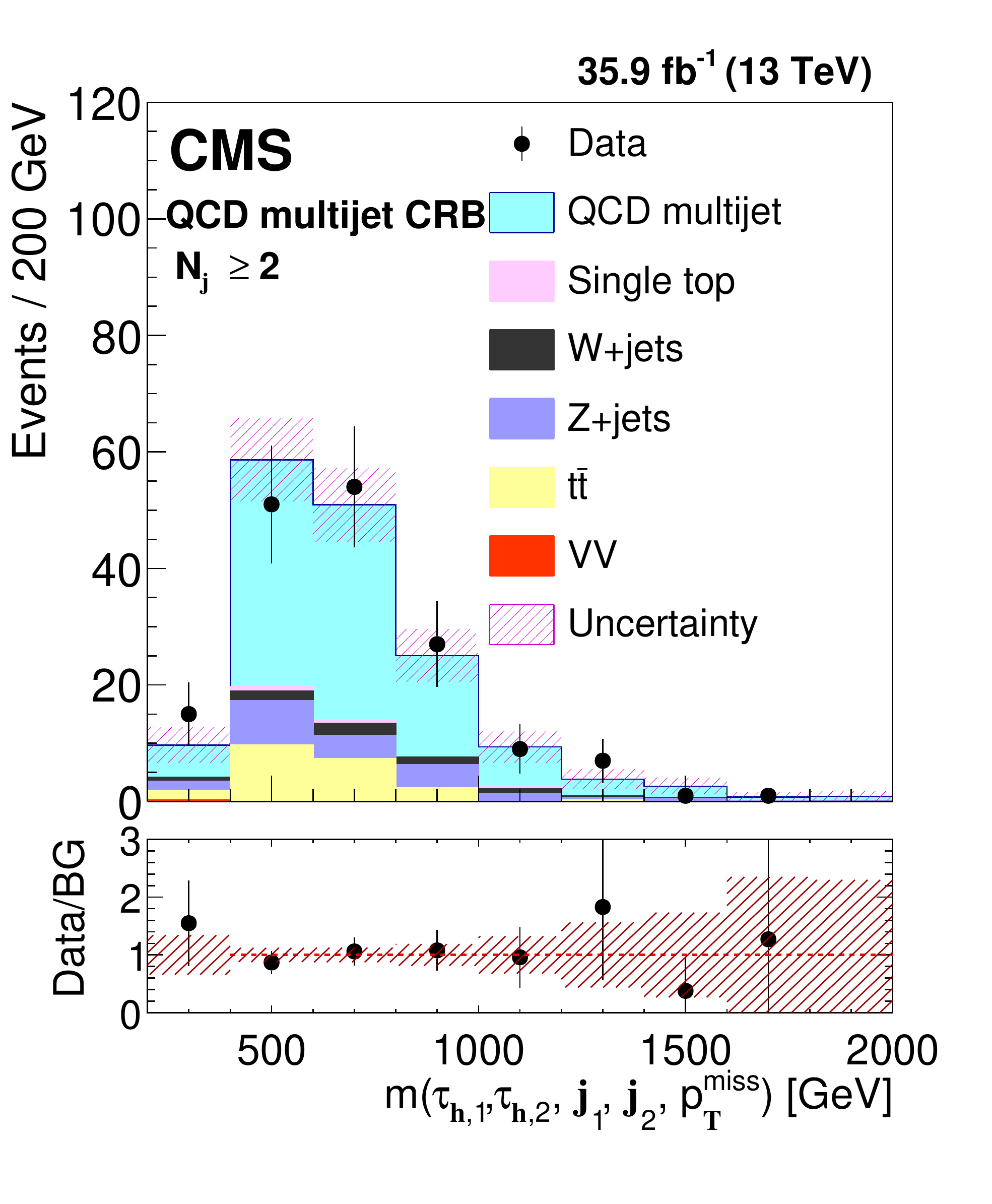}
  \includegraphics[width=0.48\textwidth, height=0.35\textheight]{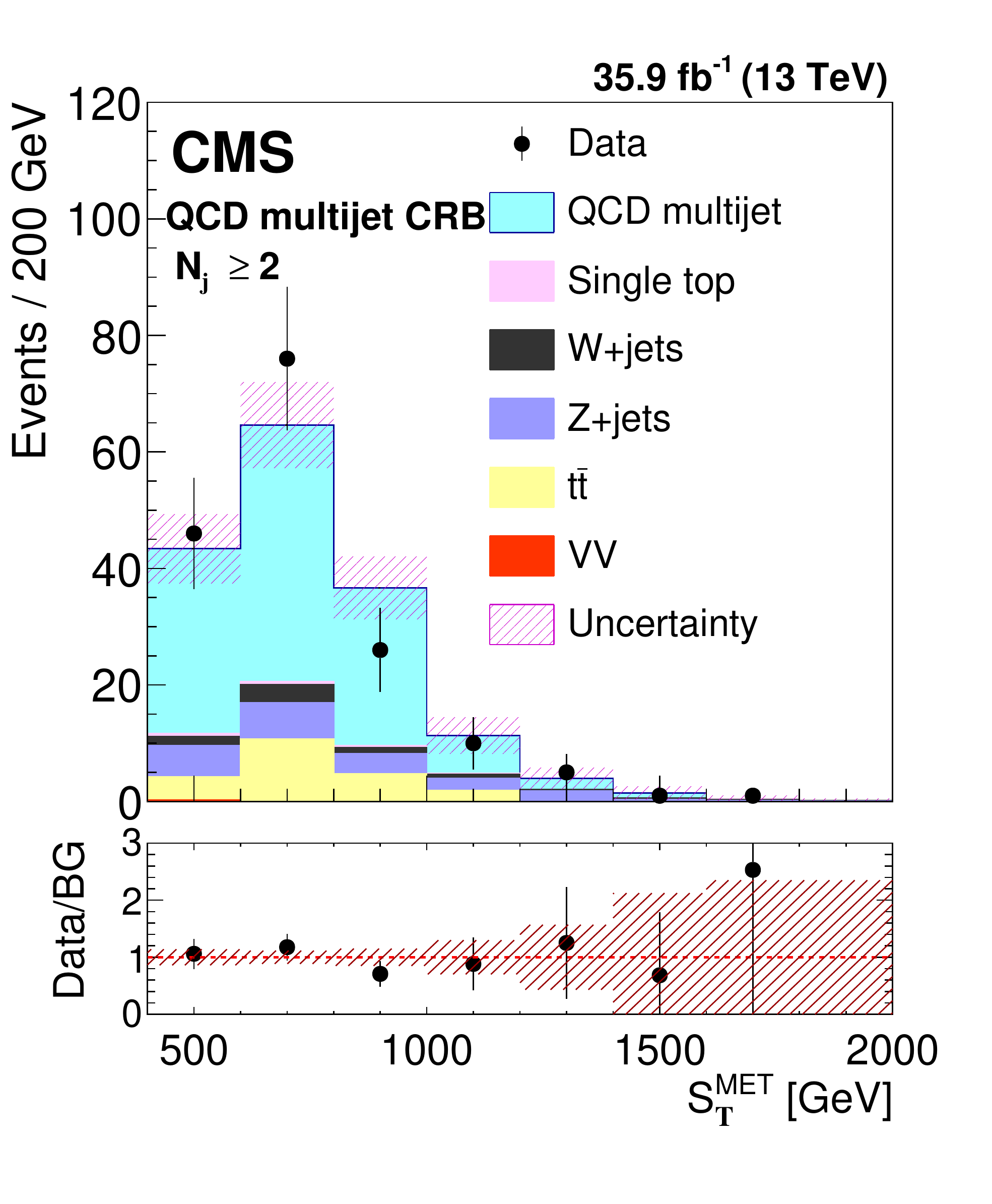}
  \caption{\label{fig:MetVsTauIso2} QCD multijet background validation test, using the distributions in CR $B$
$m(\tau_{\mathrm{h},1},\tau_{\mathrm{h},2},\mathrm{j}_{1},\mathrm{j}_{2},\ptmiss)$ (left) and $S_{\mathrm{T}}^{\mathrm{MET}}$ (right).
The shape of the QCD background is found from data in the loose $\tauh$ region, CR $A$ and then applied to CR $B$, defined by
$\pt^{miss} < 50\GeV$ and tight $\tauh$ isolation.  For both samples, the non-QCD contributions are estimated from simulation.
Note that the normalizations match by construction. The bottom frame shows the ratio between the
observed data in CR $B$ and the total background estimation.}
\end{figure}

To check that the shapes of the
$m(\tau_{\mathrm{h},1},\tau_{\mathrm{h},2},\mathrm{j}_{1},\mathrm{j}_{2},\ptmiss)$ and $S_{\mathrm{T}}^{\mathrm{MET}}$ distributions
obtained from the nonisolated CR are the same as the ones in the isolated region, we use events from
QCD-enriched CRs $A$ and $B$. Figure~\ref{fig:MetVsTauIso2} shows the $m(\tau_{\mathrm{h},1},\tau_{\mathrm{h},2},\mathrm{j}_{1},\mathrm{j}_{2},\ptmiss)$ and
$S_{\mathrm{T}}^{\mathrm{MET}}$ distributions in CR $B$. The shape of QCD multijet events is obtained from data in CR $A$, after subtracting non-QCD contributions using the simulation.
The expected QCD multijet yield is calculated as $N_{\mathrm{QCD}}^{B} = N_{\text{Data}}^{B}-N_{\neq \mathrm{QCD}}^{B}$, such
that the total background yield matches the observed number of events in data.
Therefore, the focus of this test is the overall agreement
of the QCD multijet shapes extracted from the nonisolated $\tauh$ region, as applied to the isolated region.
The agreement between the data and the predicted background distributions in Fig.\,\ref{fig:MetVsTauIso2} gives confidence that the
$m(\tau_{\mathrm{h},1},\tau_{\mathrm{h},2},\mathrm{j}_{1},\mathrm{j}_{2},\ptmiss)$ and $S_{\mathrm{T}}^{\mathrm{MET}}$
shapes for the QCD multijet background can be extracted from the nonisolated side-band and helps reduce the uncertainty in the final QCD multijet background estimate.

\section{Systematic uncertainties}\label{sec:systematics}

The imperfect MC modeling of the background processes considered in this analysis can affect the normalizations and shapes of the
$m(\tau_{\mathrm{h},1},\tau_{\mathrm{h},2},\mathrm{j}_{1},\mathrm{j}_{2},\ptmiss)$ and $S^{\mathrm{MET}}_{\mathrm{T}}$ distributions
used for the final result. Therefore, these effects are included as systematic uncertainties.
The following systematic uncertainties are considered. A \pt-dependent uncertainty
per $\tauh$ candidate in the measured trigger efficiency results in a 6\% uncertainty in the signal and background predictions
that rely on simulation.
The trigger efficiency is measured per $\tauh$ candidate by calculating the fraction of
$\PZ(\to\tau\tau\to \mu\tauh)$ events (selected with a single-$\mu$ trigger), that also pass a $\mu$-$\tauh$
trigger that has the same $\tauh$ trigger requirements as the $\tauh\tauh$ trigger used to define the SR.
Systematic effects related to the correct $\tauh$ identification are measured to be 5\% per $\tauh$
candidate\,\cite{Zxsection}. This effect is estimated from a fit to the $\PZ(\to\tau\tau)$ visible mass distribution, using the
production cross section measured in the $\PZ(\to \Pe\Pe)$ and $\PZ(\to \mu\mu)$ final states.
An additional asymmetric systematic uncertainty of $+5\%$ and $-35\%$ at $\pt = 1\TeV]$~\cite{Sirunyan:2018pgf} that increases linearly with \pt is included
to account for the extrapolation in the $\tauh$ identification efficiency estimate, which is mostly determined by low-\pt
hadronic $\tau$ lepton decays close to the $\PZ$ boson peak, to the higher-\pt regimes relevant to this analysis.
A 3\% uncertainty in the reconstructed $\tauh$ energy scale (TES) is used to assign a systematic uncertainty in both the predicted
yields and the mass and $S_{\mathrm{T}}^{\mathrm{MET}}$ shapes for signal and background with total or partial MC estimation~\cite{Sirunyan:2018pgf}. This effect ranges from
3 to 9\% depending on the sample. Systematic effects on normalization and shapes due to the uncertainty in the jet energy
scale (JES) (2--5\% depending on \pt and $\eta$) are also included, resulting in 5 to 9\% uncertainty in the normalization, depending on the sample.
Systematic uncertainties in the shapes, based on the level of agreement between the data and MC distributions in the
control samples, are also assigned. The data-to-simulation ratios of the mass and $S_{\mathrm{T}}^{\mathrm{MET}}$ distributions are fit with a first-order polynomial.
The deviation of the fit from unity, as a function of mass or $S_{\mathrm{T}}^{\mathrm{MET}}$, is assigned as a systematic uncertainty in the shape.
This results in up to $20\%$ systematic uncertainty in a given bin. 
We have checked that the choice of a first-order polynomial for the fit function adequately describes potential differences between data and MC simulation. 
A 2.5\% uncertainty comes from the measurement of the total integrated luminosity~\cite{CMS-PAS-LUM-17-001},
and affects signal and all backgrounds that are determined (in part or entirely) by simulation.

Other contributions to the total systematic uncertainty in the predicted background yields arise from the validation tests and from the statistical uncertainties associated with the data control
regions used to determine the $SF^{\ttbar}$, $SF^{\PZ\to\mu\mu}_{\text{dijet}}$, and TL factors.
The relative systematic uncertainties in $SF^{\ttbar}$ and $SF^{\PZ\to\mu\mu}_{\text{dijet}}$ related to the statistical precision in the CRs range
between 1 and 2\%, depending on the background component. For the QCD multijet background, the systematic uncertainty is dominated by the statistical uncertainty in the TL factor (18\%).
The systematic uncertainties in the $SF^{\ttbar}$, $SF^{\PZ\to\mu\mu}_{\text{dijet}}$, and TL
factors, evaluated from the validation tests with data and from the subtraction of nontargeted backgrounds, range from 3\% for $SF^{\ttbar}$ to 10\% for TL.

The uncertainty in the signal acceptance (6\%) associated with the choice of the PDF set included in the simulated samples is evaluated in accordance to the PDF4LHC
recommendation~\cite{PDF4LHCRunII, Alekhin:2011sk, Botje:2011sn}.
The absence of higher-order contributions to the cross sections affect the signal acceptance calculation.
This effect is estimated by varying the renormalization and factorization scales a factor of two
with respect to their nominal values, and by considering the full change in the yields.
They are estimated from simulation and found to be small for both signal ($2.5$\%) and background (1\% for diboson and 3.5\% for
\ttbar).
Table \ref{table:SystematicsTable} summarizes the systematic uncertainties considered in the analysis. The total systematic uncertainties
in the background normalizations range from 18 to 37\%, depending on the background, while the total systematic uncertainty in the signal
normalization is approximately 15\%.

\begin{table}\centering
 \topcaption{Summary of systematic uncertainties, given in
   percent. The $\tauh$ identification, JES, and TES uncertainties are also considered as uncertainties in the shapes of the
$m(\tau_{\mathrm{h},1},\tau_{\mathrm{h},2},\mathrm{j}_{1},\mathrm{j}_{2},\ptmiss)$ and $S^{\mathrm{MET}}_{\mathrm{T}}$ distributions.
Not included in the table are the bin-by-bin statistical uncertainties, which increase with larger values of mass and $S_{\mathrm{T}}^{\mathrm{MET}}$.}
 \begin{tabular}{lcccccc}  \hline
   Source                           & QCD        & {\PW}+jets     & {\PZ}+jets    & \ttbar  & $\mathrm{VV}$ & Signal      \\
   \hline Integrated luminosity     & \NA         & 2.5   & 2.5   & 2.5     & 2.5     & 2.5     \\
    $\tauh\tauh$ trigger  & \NA         & 6     & 6     & 6       & 6       & 6 \\
    $\tauh$ identification    & \NA         & 33  & 10  & 10    & 12    & 10  \\
    JES                       & \NA         & 9   & 8   & 6     & 9     & 5   \\
    TES                       & \NA         & 9   & 9   & 9     & 8     & 3   \\
    PDF                       & \NA         & 6     & 6     & 6       & 6       & 6     \\
    Scales                     & \NA         & 1     & 1     & 3.5     & \NA      & 2.5   \\
    Background est.: closure+norm.    & 21         & \NA    & 7     & 3       & \NA      & \NA    \\
   \hline
 \end{tabular}
 \label{table:SystematicsTable}
\end{table}

\section{Results}\label{sec:Results}

\begin{figure}[ht]
  \centering
    \includegraphics[width=0.48\textwidth, height=0.35\textheight]{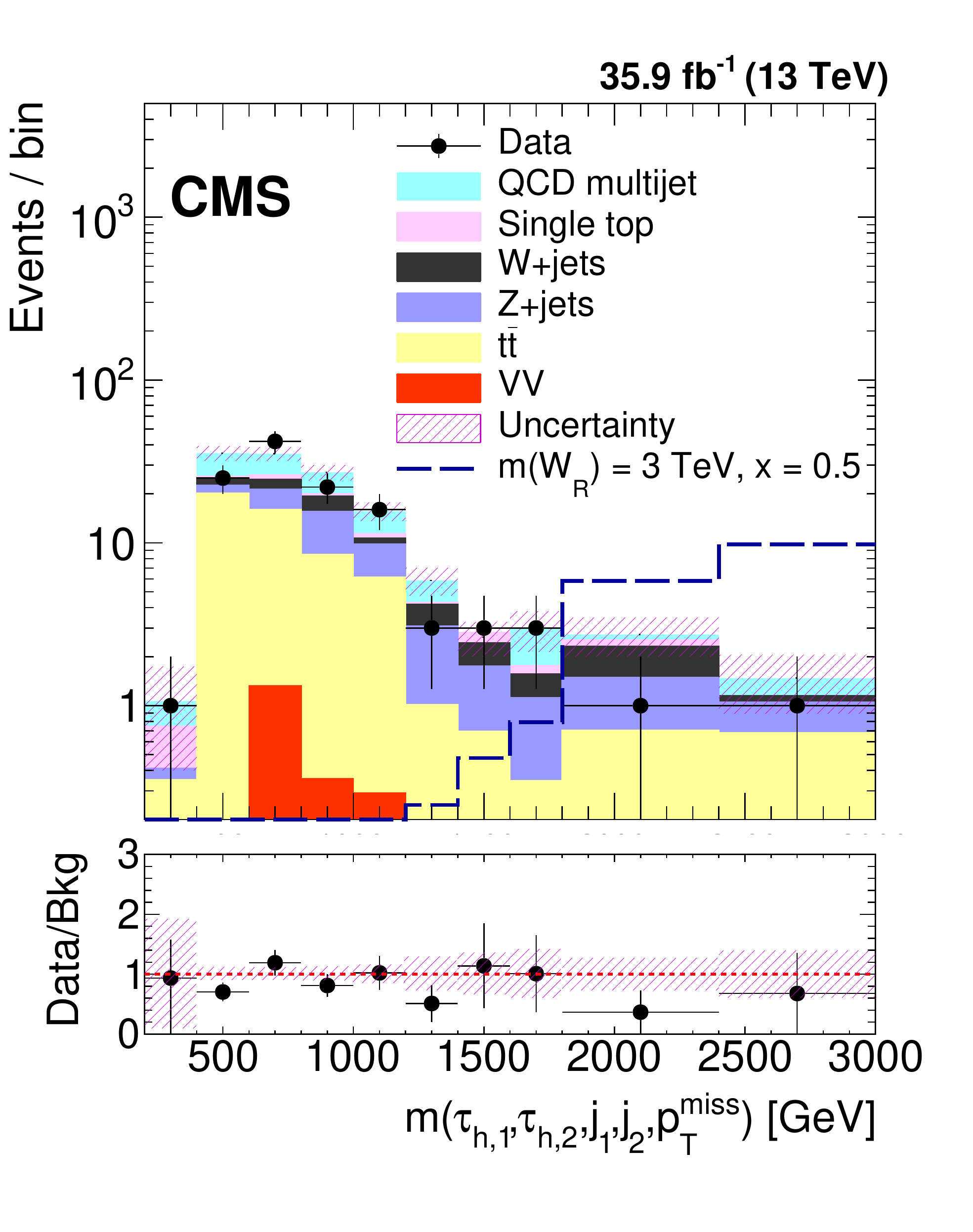}
    \includegraphics[width=0.48\textwidth, height=0.35\textheight]{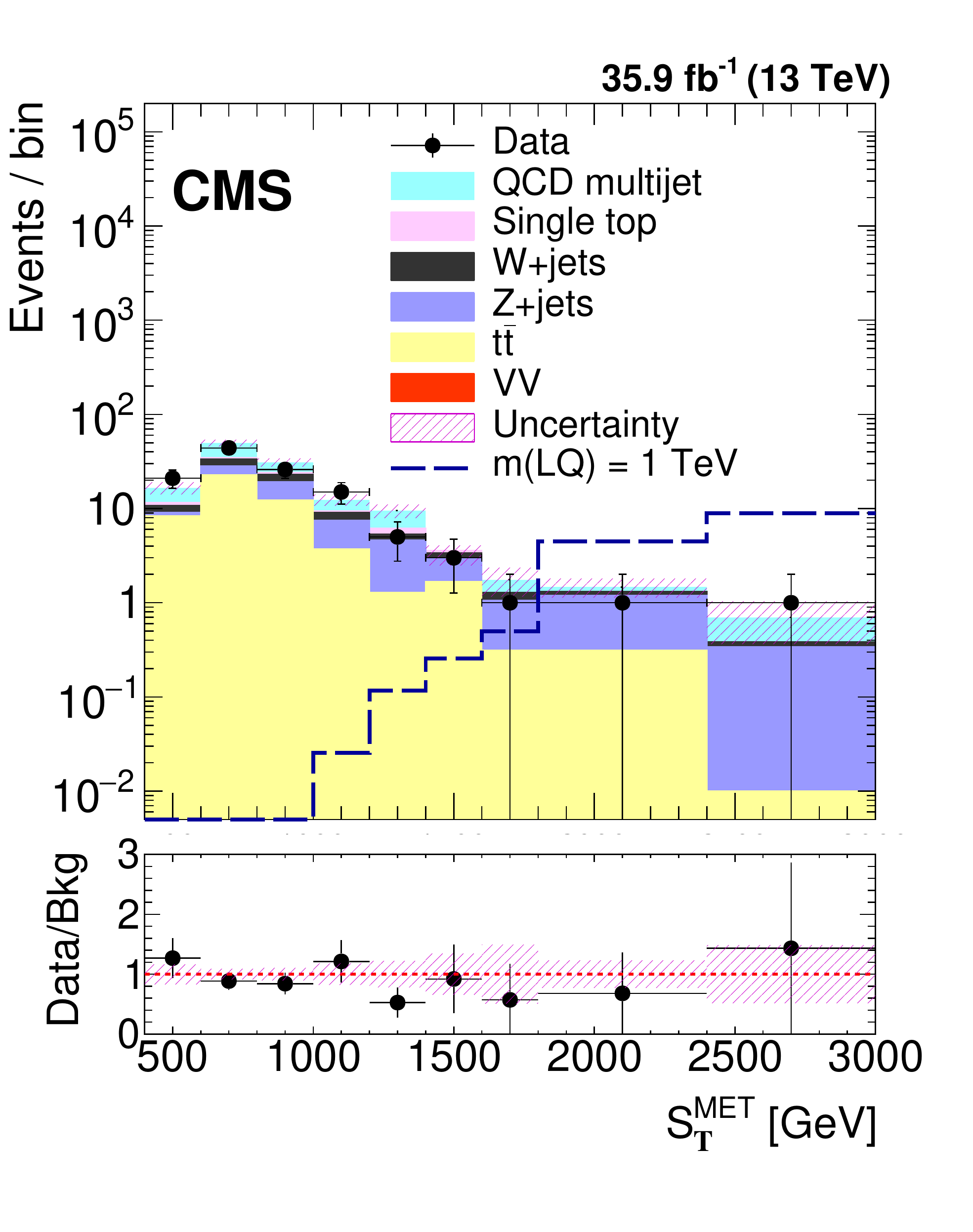}
  \caption{ Distributions in $m(\tau_{\mathrm{h},1},\tau_{\mathrm{h},2},\mathrm{j}_{1},\mathrm{j}_{2},\ptmiss)$ (left) and $S_{T}^{\mathrm{MET}}$ (right)
for the estimated background in the signal region.
The heavy neutrino model with $m(\PWR) = 3\TeV$ and $m(\Nt) = 1.5\TeV$ is used as a benchmark in the
$m(\tau_{\mathrm{h},1},\tau_{\mathrm{h},2},\mathrm{j}_{1},\mathrm{j}_{2},\ptmiss)$ distribution,
while the leptoquark model with $m(\LQ) = 1\TeV$ is used as a benchmark in the $S_{T}^{\mathrm{MET}}$ distribution.
The bottom frame shows the ratio between the observed data and the
background estimation; the band corresponds to the statistical uncertainty in the background. The \ttbar,
QCD multijet, and {\PZ}+jets contributions are estimated employing control regions in data and simulation, while the other contributions
are obtained fully from the simulation.}
  \label{fig:SignalRegionPlot}
  \end{figure}

The observed yield is 117 events, while the total predicted background yield is $127.0 \pm 11.8$ events (see Table~\ref{tab:totalBGpre}).
Table~\ref{tab:totalBGpre} illustrates the relative importance of the different backgrounds. Note, however, that the relative yields
of different background processes do not directly reflect the effect on the sensitivity of the analysis, as a binned maximum likelihood fit,
in which shape information enters besides the yields, is used to set limits on the signal rate.
Figure\,\ref{fig:SignalRegionPlot} shows the background predictions, the observed data, and the expected signal in the
$m(\tau_{\mathrm{h},1},\tau_{\mathrm{h},2},\mathrm{j}_{1},\mathrm{j}_{2},\ptmiss)$ and $S^{\mathrm{MET}}_{\mathrm{T}}$ distributions.
The heavy neutrino model with $m(\PWR) = 3.0\TeV$ and $m(\Nt) = 1.5\TeV$ is used as a benchmark in Fig.\,\ref{fig:SignalRegionPlot} (left),
while the leptoquark model with $m(\LQ) = 1.0$\TeV is used as a benchmark in Fig.\,\ref{fig:SignalRegionPlot} (right).
The observed data event rate and shapes are consistent with the SM background expectation. Therefore, exclusion limits for the
two signal benchmark scenarios are set, using the distribution in
$m(\tau_{\mathrm{h},1},\tau_{\mathrm{h},2},\mathrm{j}_{1},\mathrm{j}_{2},\ptmiss)$
for the $\Nt$ case and in $S^{\mathrm{MET}}_{\mathrm{T}}$ for the \LQ interpretation.
The results are presented as 95\% confidence level (\CL) upper limits on the signal production cross sections,
estimated with the modified frequentist construction \CLs method~\cite{Junk,CLs,LHC-HCG}.
Maximum likelihood fits are performed using the final $m(\tau_{\mathrm{h},1},\tau_{\mathrm{h},2},\mathrm{j}_{1},\mathrm{j}_{2},\ptmiss)$ and $S^{\mathrm{MET}}_{\mathrm{T}}$
discrimination
variables to derive the expected and observed limits.
Systematic uncertainties are represented by nuisance parameters, assuming a gamma function prior for the uncertainties in the data-driven background
estimations, log-normal prior for MC-driven normalization parameters, and Gaussian priors for the shape uncertainties.
Statistical uncertainties in the shape templates are accounted for by the technique described in Ref.~\cite{Barlow:1993dm}.

\begin{figure}
  \centering
    \includegraphics[width=0.49\textwidth, height=0.35\textheight]{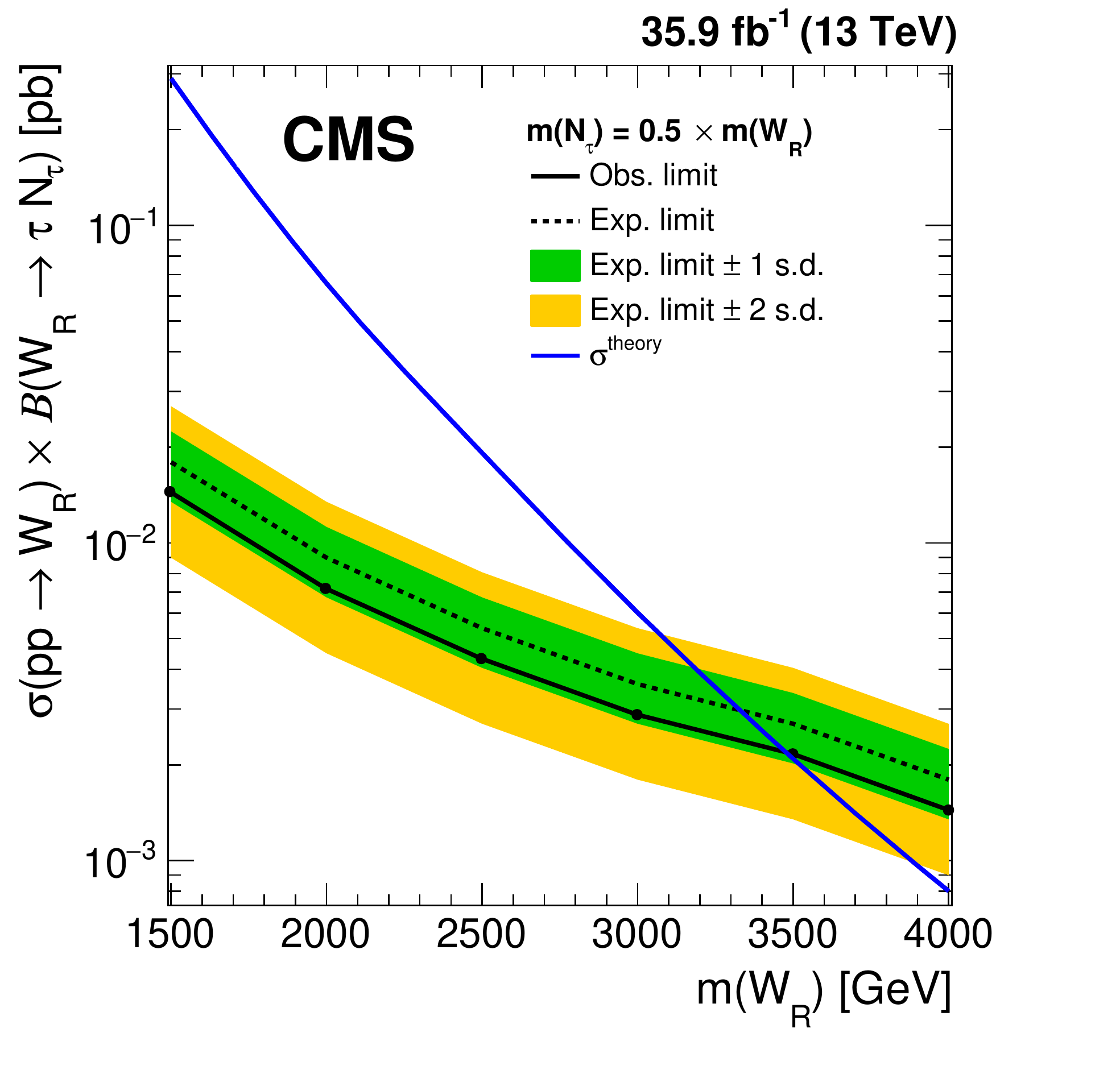}
    \includegraphics[width=0.49\textwidth, height=0.35\textheight]{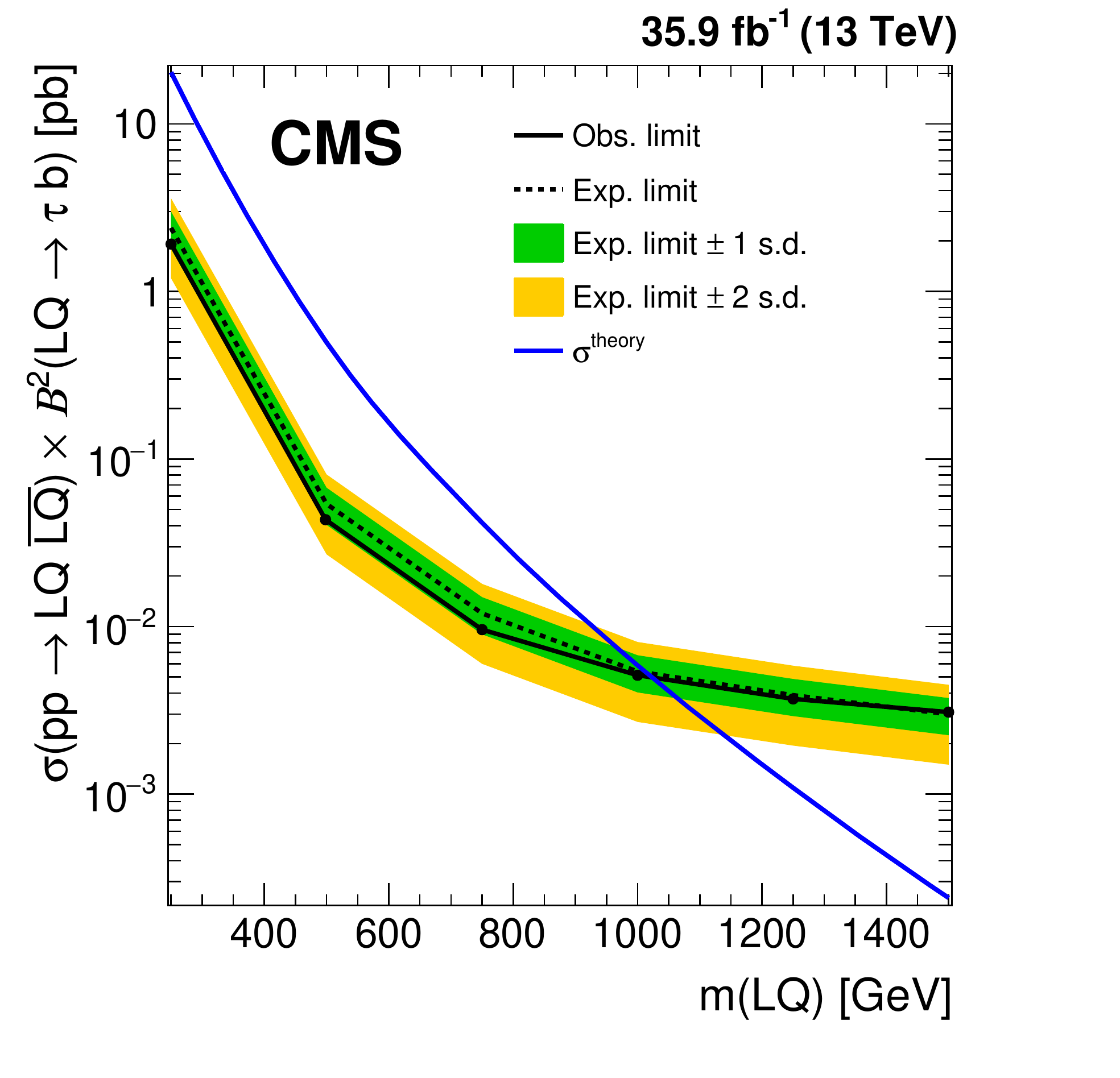}
  \caption{Upper limits at 95\% \CL on the product of the cross section and the branching fraction for the production of
$\PWR$ (left) decaying to $\Nt$ and for a pair of leptoquarks each decaying to $\tau \PQb$ (right),
as functions of the produced particle mass. The observed limits are shown as solid black lines.
Expected limits and their one- (two-) standard deviation limits are shown by dashed lines with green (yellow) bands. The theoretical cross sections are indicated by the solid blue lines.}
    \label{fig:OneDLimits}
\end{figure}

Figure\,\ref{fig:OneDLimits} shows the expected and observed limits on the cross section, as well as the theoretical prediction~\cite{Kramer:2004df,delAguila:2007qnc},
as functions of $m(\PWR)$ and $m(\LQ)$. For heavy neutrino models with strict left-right symmetry,
with the assumptions that only the $\Nt$ flavor contributes significantly to the $\PWR$ decay width and that
the $\Nt$ mass is $0.5 \times m(\PWR)$, $\PWR$ masses below 3.50\TeV are excluded at 95\% \CL (expected exclusion  3.35\TeV).
For the \LQ interpretation using $S^{\mathrm{MET}}_{\mathrm{T}}$ as the final fit variable, the observed (expected) 95\% \CL exclusion is
1.02 (1.00)\TeV. These results are the most stringent limits to date.

\begin{figure}
  \centering
    \includegraphics[width=0.7\textwidth]{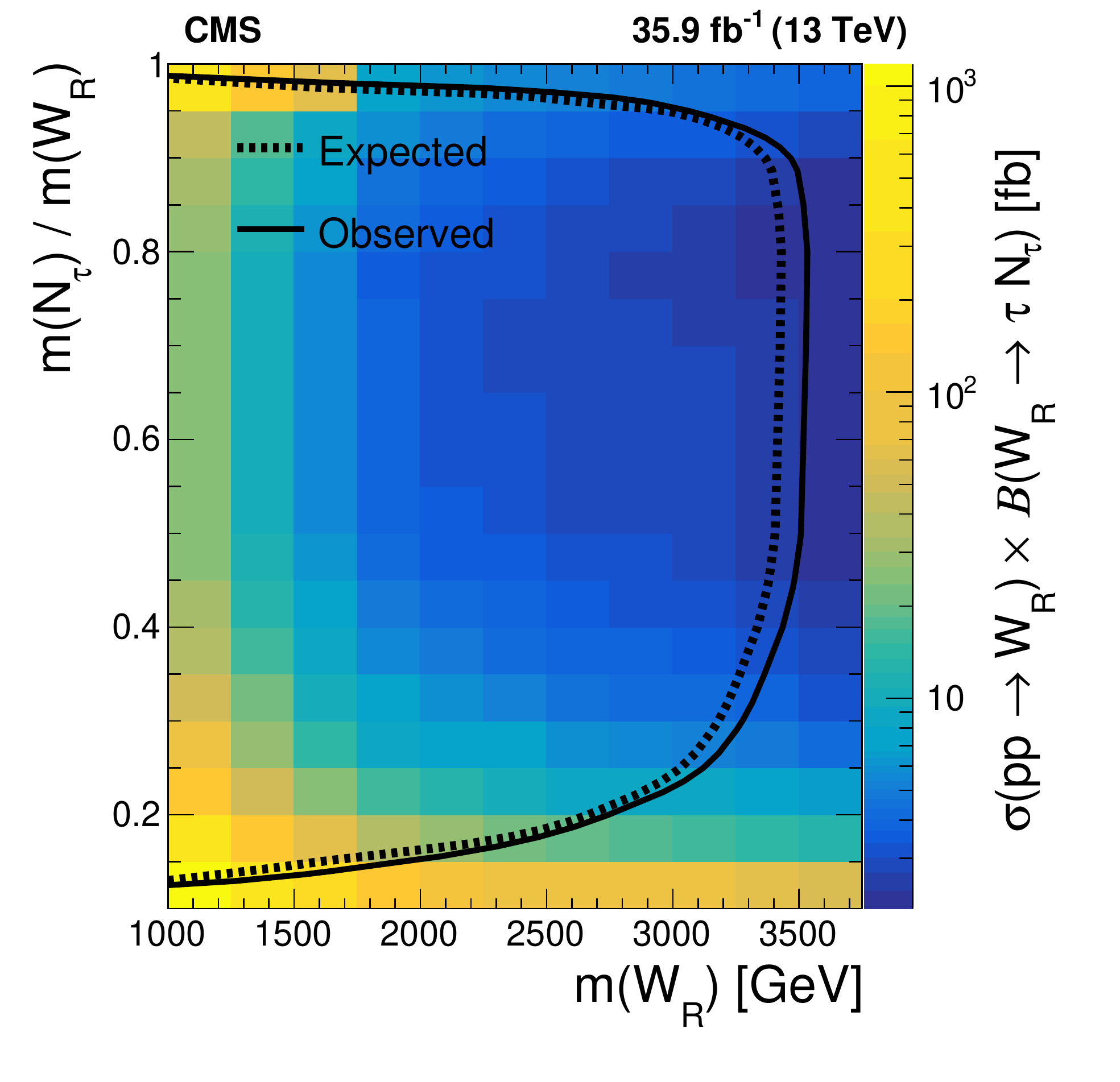}
  \caption{Expected and observed limits at 95\% \CL on the product of the cross section and the branching fraction
($\PWR \to \tau \Nt$) as a function of $m(\PWR)$ and $m(\Nt)/m(\PWR)$.}
    \label{fig:2Dlimit}
\end{figure}

Figure~\ref{fig:2Dlimit} shows 95\% \CL upper limits on the product of the production cross section and branching fraction,
as a function of $m(\PWR)$ and $x = m(\Nt)/m(\PWR)$. The signal acceptance and mass shape are evaluated for
each \{$m(\PWR), x$\} combination and used in the limit calculation procedure described above.
The $\PWR$ limits depend on the $\Nt$ mass.
For example, a scenario with $x = 0.1$ (0.25) yields significantly lower average
jet and subleading $\tauh$ \pt than the $x = 0.5$ mass assumption, and the acceptance is lower
by a factor of about 16 (3) for
$m(\PWR) = 1.0\TeV$ and about 5.8 (1.8) for $m(\PWR) = 3.0\TeV$.
On the other hand, the $x = 0.75$ scenario produces similar or larger average \pt for the jet and the $\tau_{h}$
than the $x = 0.5$ mass assumption, yielding an event acceptance that is about 10\% larger.
Masses below $m(\PWR)=3.52$ (2.75)\TeV are excluded at 95\% \CL, assuming that the $\Nt$ mass is 0.8 (0.2) times
the mass of the $\PWR$ boson.

\begin{table}
 \topcaption{Estimated background and signal yields in the SR and their total uncertainties.
The expected number of events for the $\PWR$ signal sample assumes $m(\Nt) = m(\PWR)/2$.}
  \centering{
    \begin{tabular}{ l  c  }
      \hline
Process & Yield \\ \hline
\ttbar  & $49.8 \pm 11.8$  \\
QCD             & $33.8 \pm 9.3$ \\
{\PZ}+jets & $23.4 \pm 6.5$ \\
{\PW}+jets & $13.4 \pm 6.2$ \\
Single top      & $4.6 \pm 2.2$\\
VV              & $2.0 \pm 1.5$ \\ [\cmsTabSkip]
Total           & $127.0 \pm 17.7$\\[\cmsTabSkip]
Observed        & $117$                \\[\cmsTabSkip]
$m(\PWR)=3.0$\TeV & $17.3 \pm 2.5$    \\
$m(\LQ)=1.0$\TeV & $14.2 \pm 2.1$    \\
      \hline
    \end{tabular}
  }
  \label{tab:totalBGpre}
\end{table}

\section{Summary}\label{sec:Summary}

A search is performed for physics beyond the standard model in events with two energetic $\tau$ leptons and two energetic jets, using
data corresponding to an integrated luminosity of 35.9\fbinv collected in 2016 with the CMS detector in proton-proton collisions at $\sqrt{s}=13\TeV$.
The search focuses on two benchmark scenarios: (1) the production of heavy right-handed Majorana neutrinos, $\Nell$, and right-handed
$\PWR$ bosons, which arise in the left-right symmetric extensions of the standard model and where the $\PWR$ and $\Nell$ decay chains result in a pair of
high transverse momentum $\tau$ leptons; and (2) the pair production of third-generation scalar leptoquarks that decay to $\tau\tau\PQb\PQb$.
The observed $m(\tau_{\mathrm{h},1},\tau_{\mathrm{h},2},\mathrm{j}_{1},\mathrm{j}_{2},\ptmiss)$ and $S^{\mathrm{MET}}_{\mathrm{T}}$ distributions do not reveal any evidence
for physics beyond the standard model.
Assuming that only the $\Nt$ flavor contributes significantly to the $\PWR$ decay width, $\PWR$
masses below 3.52\,(2.75)\TeV are excluded at 95\% confidence level, assuming the $\Nt$ mass is 0.8\,(0.2) times the mass of the $\PWR$ boson.
In the second beyond the standard model scenario, leptoquarks with a mass less than 1.02\TeV are excluded at 95\% confidence level, to be compared
with an expected mass limit of 1.00\TeV. Both of these results represent the most stringent limits to date for $\tau \tau \mathrm{j} \mathrm{j}$ final states.

\begin{acknowledgments}
We congratulate our colleagues in the CERN accelerator departments for the excellent performance of the LHC and thank the technical and administrative staffs at CERN and at other CMS institutes for their contributions to the success of the CMS effort. In addition, we gratefully acknowledge the computing centres and personnel of the Worldwide LHC Computing Grid for delivering so effectively the computing infrastructure essential to our analyses. Finally, we acknowledge the enduring support for the construction and operation of the LHC and the CMS detector provided by the following funding agencies: BMBWF and FWF (Austria); FNRS and FWO (Belgium); CNPq, CAPES, FAPERJ, FAPERGS, and FAPESP (Brazil); MES (Bulgaria); CERN; CAS, MoST, and NSFC (China); COLCIENCIAS (Colombia); MSES and CSF (Croatia); RPF (Cyprus); SENESCYT (Ecuador); MoER, ERC IUT, and ERDF (Estonia); Academy of Finland, MEC, and HIP (Finland); CEA and CNRS/IN2P3 (France); BMBF, DFG, and HGF (Germany); GSRT (Greece); NKFIA (Hungary); DAE and DST (India); IPM (Iran); SFI (Ireland); INFN (Italy); MSIP and NRF (Republic of Korea); MES (Latvia); LAS (Lithuania); MOE and UM (Malaysia); BUAP, CINVESTAV, CONACYT, LNS, SEP, and UASLP-FAI (Mexico); MOS (Montenegro); MBIE (New Zealand); PAEC (Pakistan); MSHE and NSC (Poland); FCT (Portugal); JINR (Dubna); MON, RosAtom, RAS, RFBR, and NRC KI (Russia); MESTD (Serbia); SEIDI, CPAN, PCTI, and FEDER (Spain); MOSTR (Sri Lanka); Swiss Funding Agencies (Switzerland); MST (Taipei); ThEPCenter, IPST, STAR, and NSTDA (Thailand); TUBITAK and TAEK (Turkey); NASU and SFFR (Ukraine); STFC (United Kingdom); DOE and NSF (USA).

\hyphenation{Rachada-pisek} Individuals have received support from the Marie-Curie programme and the European Research Council and Horizon 2020 Grant, contract No. 675440 (European Union); the Leventis Foundation; the A. P. Sloan Foundation; the Alexander von Humboldt Foundation; the Belgian Federal Science Policy Office; the Fonds pour la Formation \`a la Recherche dans l'Industrie et dans l'Agriculture (FRIA-Belgium); the Agentschap voor Innovatie door Wetenschap en Technologie (IWT-Belgium); the F.R.S.-FNRS and FWO (Belgium) under the ``Excellence of Science - EOS" - be.h project n. 30820817; the Ministry of Education, Youth and Sports (MEYS) of the Czech Republic; the Lend\"ulet (``Momentum") Programme and the J\'anos Bolyai Research Scholarship of the Hungarian Academy of Sciences, the New National Excellence Program \'UNKP, the NKFIA research grants 123842, 123959, 124845, 124850 and 125105 (Hungary); the Council of Science and Industrial Research, India; the HOMING PLUS programme of the Foundation for Polish Science, cofinanced from European Union, Regional Development Fund, the Mobility Plus programme of the Ministry of Science and Higher Education, the National Science Center (Poland), contracts Harmonia 2014/14/M/ST2/00428, Opus 2014/13/B/ST2/02543, 2014/15/B/ST2/03998, and 2015/19/B/ST2/02861, Sonata-bis 2012/07/E/ST2/01406; the National Priorities Research Program by Qatar National Research Fund; the Programa Estatal de Fomento de la Investigaci{\'o}n Cient{\'i}fica y T{\'e}cnica de Excelencia Mar\'{\i}a de Maeztu, grant MDM-2015-0509 and the Programa Severo Ochoa del Principado de Asturias; the Thalis and Aristeia programmes cofinanced by EU-ESF and the Greek NSRF; the Rachadapisek Sompot Fund for Postdoctoral Fellowship, Chulalongkorn University and the Chulalongkorn Academic into Its 2nd Century Project Advancement Project (Thailand); the Welch Foundation, contract C-1845; and the Weston Havens Foundation (USA).
\end{acknowledgments}

\bibliography{auto_generated}

\appendix

\cleardoublepage \appendix\section{The CMS Collaboration \label{app:collab}}\begin{sloppypar}\hyphenpenalty=5000\widowpenalty=500\clubpenalty=5000\vskip\cmsinstskip
\textbf{Yerevan Physics Institute, Yerevan, Armenia}\\*[0pt]
A.M.~Sirunyan, A.~Tumasyan
\vskip\cmsinstskip
\textbf{Institut f\"{u}r Hochenergiephysik, Wien, Austria}\\*[0pt]
W.~Adam, F.~Ambrogi, E.~Asilar, T.~Bergauer, J.~Brandstetter, M.~Dragicevic, J.~Er\"{o}, A.~Escalante~Del~Valle, M.~Flechl, R.~Fr\"{u}hwirth\cmsAuthorMark{1}, V.M.~Ghete, J.~Hrubec, M.~Jeitler\cmsAuthorMark{1}, N.~Krammer, I.~Kr\"{a}tschmer, D.~Liko, T.~Madlener, I.~Mikulec, N.~Rad, H.~Rohringer, J.~Schieck\cmsAuthorMark{1}, R.~Sch\"{o}fbeck, M.~Spanring, D.~Spitzbart, A.~Taurok, W.~Waltenberger, J.~Wittmann, C.-E.~Wulz\cmsAuthorMark{1}, M.~Zarucki
% ...
\vskip\cmsinstskip
% ...
1:  Also at Vienna University of Technology, Vienna, Austria\\
% ...
\end{sloppypar}
\end{document}